# Autocorrelation-Run Formula for Binary Sequences *

Kai Cai[†,‡]

† Institute of Computing Technology, Chinese Academy of Sciences,
Beijing 100080, P.R. China

‡ School of Mathematical Sciences, Peking University, Beijing 100871, P.R. China

E-mail: caikai@ict.ac.cn

**Abstract**

The autocorrelation function and the run structure are two basic notions for binary sequences, and have been used as two independent postulates to test randomness of binary sequences ever since Golomb 1955. In this paper, we prove for binary sequence that the autocorrelation function is in fact completely determined by its run structure.

# 1 Introduction

Let $s = (s_0, s_1, ..., s_{N-1})$ be a binary sequence of period $N$, where $s_i \in \{0, 1\}$. The *Hamming weight* of $s$ is defined as the number of 1s in a period of $s$, and denoted by $wt(s)$. The *autocorrelation function* of an $N$-period sequence $s$ can be defined as:

$$C_s(w) = N - 2wt(s \oplus T^w s), \qquad (1.1)$$

for $w = 0, 1, ..., N - 1$, where the symbol $\oplus$ denotes the addition of $\mathbb{F}_2^N$ and $T$ denotes the *left shift transformation* of a binary sequence. $C_s(1), C_s(2), \cdots, C_s(N-1)$ are called the *out-of-phase autocorrelation coefficients* of $s$. The autocorrelation function is an important characterization for binary sequences, and the sequences with lower out-of-phase autocorrelation coefficients are widely used in engineering applications [1, 2, 3]. A *run* of a binary sequence is defined as a string of consecutive identical symbols flanked at either end by the other symbol. A run with consecutive 1s is called a 1-*run* (or *block*) and a run with consecutive 0s is called a *0-run* (or *gap*) of $s$. The number of the symbols in a run is called the *length of the run*. The run structure is another important characterization for binary sequences. It is well known that some sequences, e.g., m-sequences have constant out-of-phase autocorrelation coefficients $-1$ and also some kind of "perfect" run structure (see p.62 of [1]). The autocorrelation function and the run structure have been used independently to characterize the randomness of binary sequences for more than 30 years [1, 3].

---

*This research is supported by NSFC of China (No. 60872036)

In this paper, we investigate the relation between the autocorrelation function and the run structure of a binary sequence. A formula connecting these two categories is establish. It is a run series expansion formula for the autocorrelation function. The key idea of our approach is to calculate the autocorrelation function run by run instead of symbol by symbol of the definition. The approach is developed based on the technique of integer composition analysis. Our result provides a new tool to analyze binary sequences and the related natures, e.g., difference sets, circulant Hadamard matrix and etc.

The rest of this paper is organized as follows. In Section 2, we propose some basic notations, results and observations. The main result is established in Section 3. Examples and some applications are presented in Sections 4 and 5, respectively.

## 2 Preliminaries

In this section, we first introduce some basic definitions and useful results related to the integer composition and then propose some denotations and observations related to the run structure and the autocorrelation function.

### 2.1 Integer Composition

We use $\mathbb{N}$ to denote the set of positive integers $\{1, 2, 3, \cdots\}$. A *composition* (or *ordered partition*) of $n \in \mathbb{N}$ is a string of positive integers $p = (a_1, a_2, \cdots, a_s)$ such that $\sum_{i=1}^{s} a_i = n$, where $s$ is called the length of $p$ and denoted as $|p|$. It is well known that there are totally $2^{n-1}$ compositions of $n$ [4] . The set of all the compositions of $n$ is denoted as $\mathcal{P}(n)$ in this paper. For example, $\mathcal{P}(1) = \{(1)\}$, $\mathcal{P}(2) = \{(2), (1,1)\}$ and $\mathcal{P}(3) = \{(3), (1,2), (2,1), (1,1,1)\}$. The order of the elements in $\mathcal{P}(n)$ is also taken into account. For example, the 4 elements of $\mathcal{P}(3)$ are denoted as $p_0(3) = (3)$, $p_1(3) = (1,2)$, $p_2(3) = (2,1)$, and $p_3(3) = (1,1,1)$. Generally, we write $\mathcal{P}(n) = \{p_0(n), p_1(n), \cdots, p_{2^{n-1}-1}(n)\}$. $\mathcal{P}(n)$ can be generated and simultaneously ordered as follows (see Fig. 1, with the $n$th layer $\mathcal{P}(n)$).

**Proposition 2.1** *$\mathcal{P}(n)$ can be generated inductively based on the following steps.*

1. $p_0(1) = (1)$;

2. If $p_i(n-1) = (a_1, a_2, \cdots, a_s)$, then $p_{2i}(n) = (a_1 + 1, a_2, \cdots, a_s)$ and $p_{2i+1}(n) = (1, a_1, a_2, \cdots, a_s)$, for $i = 0, 1, \cdots, 2^{n-2} - 1$.

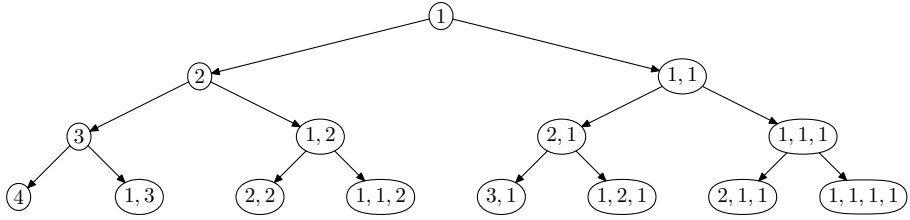

Fig. 1: Generation of the compositions.

Obviously, the above process generates all the $2^{n-1}$ different elements of $\mathcal{P}(n)$ for each $n \in \mathbb{N}$. By Proposition 2.1, for each $n \in \mathbb{N}$, we can define two mappings, namely $\varphi_I^n : \mathcal{P}(n) \longrightarrow \mathcal{P}(n+1)$ and $\varphi_{II}^n : \mathcal{P}(n) \longrightarrow \mathcal{P}(n+1)$ such that

$$\varphi_I^n : p_i(n) \longmapsto p_{2i}(n+1);$$

and

$$\varphi_{II}^n : p_i(n) \longmapsto p_{2i+1}(n+1).$$

Let $\mathcal{P}_I(n+1)$ and $\mathcal{P}_{II}(n+1)$ denote the images of $\varphi_I^n$ and $\varphi_{II}^n$, respectively. Then,

**Proposition 2.2** *The following items hold for each $n \in \mathbb{N}$.*

1. $\mathcal{P}_I(n+1) \cup \mathcal{P}_{II}(n+1) = \mathcal{P}(n+1);$

2. $\varphi_I^n$ and $\varphi_{II}^n$ are 1-1 maps from $\mathcal{P}(n)$ to $\mathcal{P}_I(n+1)$ and $\mathcal{P}_{II}(n+1)$, respectively.

For the above reason, we use $\varphi_I^{-n}$ and $\varphi_{II}^{-n}$ to denote the corresponding inverses of $\mathcal{P}_I^n$ and $\mathcal{P}_{II}^n$, which are mappings from $\mathcal{P}_I(n+1)$ and $\mathcal{P}_{II}(n+1)$ to $\mathcal{P}(n)$, respectively.

Let $p_i(n) = (a_1, a_2, \cdots, a_s) \in \mathcal{P}(n)$ and $k \in \mathbb{N}$. We use $(p_i(n), k)$ and $\{\mathcal{P}(n), k\}$ to denote $(a_1, a_2, \cdots, a_s, k)$ and $\{(p_0(n), k), (p_1(n), k), \cdots, (p_{2^{n-1}-1}(n), k)\}$, respectively. Using these denotations, we have

**Proposition 2.3** *For any $n = 2, 3, 4, \cdots$, it holds that*

$$\mathcal{P}(n) = \{(n)\} \bigcup_{\substack{1 \leq i \leq n-1 \\ i+j=n}} \{\mathcal{P}(i), j\}$$

## 2.2 Run Strings

In this paper, the total number of runs of a binary sequence is denoted as $\gamma$. Note that $\gamma$ is always an even integer. A run with length $i$ is denoted by $R_i$ (we also call the run with type $R_i$). A string of consecutive runs with lengths $i_1, i_2, \cdots i_s$ is denoted as $R_{i_1} R_{i_2} \cdots R_{i_s}$ (we also call it with type $R_{i_1} R_{i_2} \cdots R_{i_s}$). Note that each run (run string) type associates with two runs (run strings). For example, $R_1 R_2$ associates with 100 and 011. Let $A = (a_1, a_2, a_3, \cdots)$ be a finite or infinite string and $B = (b_1, b_2, \cdots b_s)$ be a finite string of positive integers. We use the following denotations.

- $R^A$: the run string $R_{a_1} R_{a_2} R_{a_3} \cdots$.
- $R^{A^+}$: the collection of the run strings $R_i R_{a_2} R_{a_3} \cdots$ for $i \geq a_1$.
- $R_A$: the collection of the runs $R_{a_1}, R_{a_2}, R_{a_3}, \cdots$.
- $R^B \ltimes R_A$: the collection of the run strings $R^B R_{a_1}, R^B R_{a_2}, R^B R_{a_3}, \cdots$.
- $R^{B^+} \ltimes R_A$: the collection of the run strings $R^{B^+} R_{a_1}, R^{B^+} R_{a_2}, R^{B^+} R_{a_3}, \cdots$.

In the above, $R^{B^+} R_{a_i}$ represents the collection of the run strings $R_j R_{b_2} R_{b_3} \cdots R_{b_s} R_{a_i}$, for $j \geq b_1$. Let $\mathcal{R}$ be a collection of run string types. We use $\mathcal{N}_s(\mathcal{R})$ to denote the number of the run strings of $s$ which have a particular type in $\mathcal{R}$. It is easy to see that $\mathcal{N}_s(\cdot)$ has the following absorptive laws:

1. $\sum_{i=1}^{\infty} \mathcal{N}_s(R_i R^A) = \mathcal{N}_s(R^A)$, for any run string $R^A$;
2. $\sum_{i=1}^{\infty} \mathcal{N}_s(R_1 R_2 \cdots R_s R_i) = \mathcal{N}_s(R_1 R_2 \cdots R_s)$.

For example, let $s = (110100000011001010111100)$, one can see that $\gamma = \sum_{i=1}^{+\infty} \mathcal{N}_s(R_i) = 12$. That is, the total number of runs within $s$ is 12, including which $\mathcal{N}_s(R_1) = 6, \mathcal{N}_s(R_2) = 4, \mathcal{N}_s(R_4) = 1$, and $\mathcal{N}_s(R_6) = 1$. One can also have $\mathcal{N}_s(R_1 R_1) = 4, \mathcal{N}_s(R_1 R_2) = 0$, $\mathcal{N}_s(R_2 R_1) = 2$ and $\mathcal{N}_s(R_1 R_1 R_1) = 2$. It can be checked that $C_s(1) = C_s(2) = C_s(3) = C_s(4) = 0$ by the definition. In fact, this can be deduced directly from the above run string numbers, which will be seen in the next section.

## 2.3 Observations

Let $s$ be a binary sequence with period $N$. It can be seen that each run of $s$ contribute 1 to $wt(s \oplus T^1 s)$, and hence $wt(s \oplus T^1 s) = \gamma$. Thus $C_s(1) = N - 2\gamma$ by Equation 1.1. The following proposition shows that $C_s(2)$ can be determined by $N$, $\gamma$, and $\mathcal{N}_s(R_1)$.

**Proposition 2.4** $C_s(2) = N - 4\gamma + 4\mathcal{N}_s(R_1)$.

**Proof** We analyze how the $\gamma$ runs contributes to $wt(s \oplus T^2 s)$. Consider the types of the runs of $s$ and distinguish them into 4 classes:

1. $R_1$, and with an $R_1$ ahead; In this case, it contributes 0 to $wt(s \oplus T^2 s)$;

2. $R_1$, and with an $R_j$ $(j \geq 2)$ ahead; In this case, it contributes 1 to $wt(s \oplus T^2 s)$;

3. $R_i$ $(i \geq 2)$, and with an $R_1$ ahead; In this case, it contributes 1 to $wt(s \oplus T^2 s)$;

4. $R_i$ $(i \geq 2)$, and with an $R_j$ $(j \geq 2)$ ahead; In this case, it contributes 2 to $wt(s \oplus T^2 s)$.

Thus, we have

$$\begin{aligned} wt(s \oplus T^2 s) &= \sum_{j \geq 2} \mathcal{N}_s(R_j R_1) + \sum_{i \geq 2} \mathcal{N}_s(R_1 R_i) + 2\sum_{j,i \geq 2} \mathcal{N}_s(R_j R_i) \\ &= \left( \sum_{j \geq 2} \mathcal{N}_s(R_j R_1) + \sum_{j,i \geq 2} \mathcal{N}_s(R_j R_i) \right) + \left( \sum_{i \geq 2} \mathcal{N}_s(R_1 R_i) + \sum_{j,i \geq 2} \mathcal{N}_s(R_j R_i) \right) \\ &= \sum_{j \geq 2} \mathcal{N}_s(R_j) + \sum_{i \geq 2} \mathcal{N}_s(R_i) \\ &= 2\gamma - 2\mathcal{N}_s(R_1) \end{aligned}$$

By Equation (1.1), we have $C_s(2) = N - 4\gamma + 4\mathcal{N}_s(R_1)$. $\square$

## 3 Main Result

The main result of this paper can be presented as three equivalent theorems.

**Theorem 3.1** *For any binary sequence $s$ with period $N$ and $t = 1, 2, 3, \cdots, N$,*

$$C_s(t) = N - 2t\gamma - 4 \sum_{i_1 + i_2 + \cdots + i_\ell < t} (-1)^\ell (t-i) \mathcal{N}_s(R_{i_1} R_{i_2} \cdots R_{i_\ell}), \quad (3.1)$$

*where $i = i_1 + i_2 + \cdots i_\ell$.*

By equation 1.1, Theorem 3.1 is equivalent to:

**Theorem 3.2** *For any binary sequence $s$ with period $N$ and $t = 1, 2, 3, \cdots, N$,*

$$wt(s \oplus T^t s) = t\gamma + 2 \sum_{i_1 + i_2 + \cdots + i_\ell < t} (-1)^\ell (t-i) \mathcal{N}_s(R_{i_1} R_{i_2} \cdots R_{i_\ell}), \quad (3.2)$$

*where $i = i_1 + i_2 + \cdots i_\ell$.*

It is easy to see that Theorem 3.2 is equivalent to:

**Theorem 3.3** *For any binary sequence $s$ with period $N$ and $t = 1, 2, 3, \cdots, N$,*

$$wt(s \oplus T^t s) - wt(s \oplus T^{t-1} s) = \gamma + 2 \sum_{i_1+i_2+\cdots+i_\ell < t} (-1)^\ell \mathcal{N}_s(R_{i_1} R_{i_2} \cdots R_{i_\ell}), \quad (3.3)$$

Note that when $t = 1$, in Equations 3.1, 3.2, and 3.3, $\mathcal{N}_s(R_{i_1} R_{i_2} \cdots R_{i_\ell}) = 0$ for all the time, and hence the theorems hold naturally. Thus, in what follows, we just prove Theorem 3.3 for the cases of $t \geq 2$. Firstly, we need some preparations.

**Definition 3.4** *Let $p_i(t) = (a_1, a_2, \cdots, a_s) \in \mathcal{P}(t)$. The function $C_{p_i(t)}(j)$ ($j \in \mathbb{N}$) is defined as follows.*

$$C_{p_i(t)}(j) = \begin{cases} 0, & \text{for } \ell \text{ odd}; \\ (-1)^{|p_i(t)|+1}, & \text{for } \ell \text{ even}, \end{cases}$$

*where $\ell = \min\{k : \sum_{i=1}^{k} a_i \geq j + 1\}$, for $j < t$ and $\ell = s + 1$ for $j \geq t$.*

**Remark 3.5** *When $t = 1$, $C_{p_0(1)}(j) \equiv 1$ for $j = 1, 2, 3, \cdots$.*

**Remark 3.6** *$C_{p_i(t)}(j)$ represent the value that $R_j$ contributes to $Wt(s \oplus T^t s) - Wt(s \oplus T^{t-1} s)$ when there is an element of $R^{p_i^+(t)}$ ahead of $R_j$.*

**Definition 3.7** *The dual of $p_i(t)$ is a subset of $\mathbb{N}$ and defined as $Q_i(t) = \{j : C_{p_i(t)}(j) \neq 0\}$.*

**Theorem 3.8** *For any $t = 2, 3, 4, \cdots$, $k = 0, 1, \cdots, t - 2$ and $j = 0, 1, \cdots, 2^{t-k-2} - 1$, it holds that*

$$Q_{2^{k+1} \cdot j}(t) \cup Q_{2^{k+1} \cdot j + 2^k}(t) = \{k+1, k+2, \cdots\}. \quad (3.4)$$

**Proof** Let $\varphi = \varphi_I^{-(t-k)} \circ \cdots \circ \varphi_I^{-(t-2)} \circ \varphi_I^{-(t-1)}$. By the arguments of Proposition 2.2,

$$\varphi(p_{2^{k+1} \cdot j}(t)) = p_{2j}(t - k);$$

$$\varphi(p_{2^{k+1} \cdot j + 2^k}(t)) = p_{2j+1}(t - k).$$

Let $p_j(t - k - 1) = (a_1, a_2, \cdots, a_s)$. We have,

$$p_{2j}(t - k) = (a_1 + 1, a_2, \cdots, a_s);$$

and

$$p_{2j+1}(t - k) = (1, a_1, a_2, \cdots, a_s).$$

Thus,

$$p_{2^{k+1} \cdot j}(t) = \varphi^{-1}(p_{2j}(t-k)) = (a_1 + k + 1, a_2, \cdots, a_s);$$
$$p_{2^{k+1} \cdot j + 2^k}(t) = \varphi^{-1}(p_{2j+1}(t-k)) = (k+1, a_1, a_2, \cdots, a_s).$$

It is easy to check that $C_{(a_1+k+1,a_2,\cdots,a_s)}(j) = C_{(k+1,a_1,a_2,\cdots,a_s)}(j) = 0$ for $j \leq k$ (The values of $\ell$ in Definition 3.4 corresponding to $(a_1+k+1, a_2, \cdots, a_s)$ and $(a_1+k+1, a_2, \cdots, a_s)$ are equal to 1 in both cases), and $C_{(a_1+k+1,a_2,\cdots,a_s)}(j) + C_{(k+1,a_1,a_2,\cdots,a_s)}(j) = \pm 1$ for each $j \geq k+1$ (The values of $\ell$ corresponding to $(a_1+k+1, a_2, \cdots, a_s)$ and $(a_1+k+1, a_2, \cdots, a_s)$ are with difference 1.). Hence,

$$Q_{2^{k+1}\cdot j}(t) \cup Q_{2^{k+1}\cdot j + 2^k}(t) = \{k+1, k+2, \cdots\}.$$

The theorem is proved. □

**Corollary 3.9** *For any $t = 2, 3, 4, \cdots$ and $j = 0, 1, 2, \cdots, 2^{t-2} - 1$,*

$$Q_{2j}(t) \cup Q_{2j+1}(t) = \mathbb{N}.$$

**Remark 3.10** $Q_0(t) = \{t, t+1, t+2, \cdots\}$, *for $t \in \mathbb{N}$.*

**Example 3.11** *The four compositions of $\mathcal{P}(3)$ are $p_0(3) = (3)$, $p_1(3) = (1, 2)$, $p_2(3) = (2, 1)$, and $p_3(3) = (1, 1, 1)$. It can be checked that $Q_0(3) = \{3, 4, 5, \cdots\}$, $Q_1(3) = \{1, 2\}$, $Q_2(3) = \{2\}$, and $Q_3(3) = \{1, 3, 4, 5, \cdots\}$ by the definition. One can see that $Q_0(3) \cup Q_1(3) = Q_2(3) \cup Q_3(3) = \mathbb{N}$ and $Q_0(3) \cup Q_2(3) = \{2, 3, 4, \cdots\}$, as Equation (3.4) indicates.*

To prove the main result, we need some denotations:

1. $\gamma_{\mathcal{P}(t)} = \sum\limits_{i=0}^{2^{t-1}-1} (-1)^{|p_i(t)|} \mathcal{N}_s(R^{p_i(t)})$.

2. $\gamma_{\{\mathcal{P}(t),k\}} = \sum\limits_{i=0}^{2^{t-1}-1} (-1)^{|p_i(t)|+1} \mathcal{N}_s(R^{(p_i(t),k)})$.

3. $\gamma_{\mathcal{P}(t)}^k = \sum\limits_{i=0}^{2^{t-k-1}-1} (-1)^{|p_i(t-k)|+1} \mathcal{N}_s(R^{p_i^+(t-k)} \ltimes R_{Q_{2^k\cdot i}(t)})$, for $t \geq 2$ and $k = 0, 1, \cdots, t-1$.

**Remark 3.12** *The following items hold for $t \geq 2$.*

1. $\gamma_{\mathcal{P}(t)} = \sum\limits_{i+j=t} \gamma_{\{\mathcal{P}(i),j\}} - \mathcal{N}_s(R_t)$.

2. $\gamma_{\mathcal{P}(t)}^{t-1} = \mathcal{N}_s(R^{(1)^+} \ltimes R_{Q_0(t)}) = \sum\limits_{i=t}^{\infty} \mathcal{N}_s(R_j) = \gamma - \sum\limits_{i=1}^{t-1} \mathcal{N}_s(R_i)$

In the above, the second equation of item 2) holds by $\mathcal{N}_s(R^{(1)^+} \ltimes R_{Q_0(t)}) = \mathcal{N}_s(R_{Q_0(t)})$ (the absorptive law of $\mathcal{N}_s(\cdot)$) and Remark 3.10. We also need the following lemma as a preparation.

**Lemma 3.13** *The following items hold for an arbitrary string of positive integers $A$ and $t \geq 2$.*

1. $\mathcal{N}_s(R^{p^+_{2i+1}(t)} \ltimes R_A) = \mathcal{N}_s(R^{p_i(t-1)} \ltimes R_A)$.

2. $\mathcal{N}_s(R^{p^+_{2i}(t)} \ltimes R_A) + \mathcal{N}_s(R^{p^+_{2i+1}(t)} \ltimes R_A) = \mathcal{N}_s(R^{p^+_i(t-1)} \ltimes R_A)$.

**Proof** *It can be seen directly from Proposition 2.1 and the absorptive law.* □

With these preparations, we can give the following theorem, which plays a central role in the proof of the main result.

**Theorem 3.14** *For any $t = 2, 3, 4, \cdots$ and $k = 1, \cdots, t-2$, it hold that*

$$\gamma^k_{\mathcal{P}(t)} = \gamma^{k+1}_{\mathcal{P}(t)} + \gamma_{\mathcal{P}(t-k-1)} + \sum_{j=1}^{k} \gamma_{\{\mathcal{P}(t-k-1),j\}}. \tag{3.5}$$

**Proof** We first rewrite $\gamma^k_{\mathcal{P}(t)}$ into two parts, namely, $A$ and $B$ for consideration.

$$\gamma^k_{\mathcal{P}(t)} = \sum_{i=0}^{2^{t-k-1}-1} (-1)^{|p_i(t-k)|+1} \mathcal{N}_s \left( R^{p^+_i(t-k)} \ltimes R_{Q_{2^k \cdot i}(t)} \right)$$

$$= \underbrace{\sum_{i=0}^{2^{t-k-2}-1} (-1)^{|p_{2i}(t-k)|+1} \mathcal{N}_s \left( R^{p^+_{2i}(t-k)} \ltimes R_{Q_{2^k \cdot 2i}(t)} \right)}_{A}$$

$$+ \underbrace{\sum_{i=0}^{2^{t-k-2}-1} (-1)^{|p_{2i+1}(t-k)|+1} \mathcal{N}_s \left( R^{p^+_{2i+1}(t-k)} \ltimes R_{Q_{2^k \cdot (2i+1)}(t)} \right)}_{B}.$$

By 2) of Lemma 3.13, we have

$$A = \sum_{i=0}^{2^{t-k-2}-1} (-1)^{|p_{2i}(t-k)|+1} \mathcal{N}_s \left( R^{p^+_{2i}(t-k)} \ltimes R_{Q_{2^{k+1} \cdot i}(t)} \right)$$

$$= \sum_{i=0}^{2^{t-k-2}-1} (-1)^{|p_{2i}(t-k)|+1} \left[ \mathcal{N}_s \left( R^{p^+_i(t-k-1)} \ltimes R_{Q_{2^{k+1} \cdot i}(t)} \right) - \mathcal{N}_s \left( R^{p^+_{2i+1}(t-k)} \ltimes R_{Q_{2^{k+1} \cdot i}(t)} \right) \right]$$

$$= \sum_{i=0}^{2^{t-k-2}-1} (-1)^{|p_{2i}(t-k)|+1} \mathcal{N}_s(R^{p^+_i(t-k-1)} \ltimes R_{Q_{2^{k+1} \cdot i}(t)})$$

$$- \sum_{i=0}^{2^{t-k-2}-1} (-1)^{|p_{2i}(t-k)|+1} \mathcal{N}_s(R^{p^+_{2i+1}(t-k)} \ltimes R_{Q_{2^{k+1} \cdot i}(t)})$$

Since $|p_i(t-k-1)| = |p_{2i}(t-k)|$ and $|p_{2i+1}(t-k)| = |p_{2i}(t-k)|+1$, we have

$$A = \underbrace{\sum_{i=0}^{2^{t-k-2}-1} (-1)^{|p_i(t-k-1)|+1} \mathcal{N}_s(R^{p_i^+(t-k-1)} \ltimes R_{Q_{2^{k+1}.i}(t)})}_{\gamma_{\mathcal{P}(t)}^{k+1}}$$
$$+ \underbrace{\sum_{i=0}^{2^{t-k-2}-1} (-1)^{|p_{2i+1}(t-k)|+1} \mathcal{N}_s(R^{p_{2i+1}^+(t-k)} \ltimes R_{Q_{2^{k+1}.i}(t)})}_{A_2}.$$

Now considering $B + A_2$, we have

$$B + A_2 = \sum_{i=0}^{2^{t-k-2}-1} (-1)^{|p_{2i+1}(t-k)|+1} \mathcal{N}_s\left(R^{p_{2i+1}^+(t-k)} \ltimes R_{Q_{2^{k+1}.i+2^k}(t)}\right)$$
$$+ \sum_{i=0}^{2^{t-k-2}-1} (-1)^{|p_{2i+1}(t-k)|+1} \mathcal{N}_s\left(R^{p_{2i+1}^+(t-k)} \ltimes R_{Q_{2^{k+1}.i}(t)}\right)$$
$$= \sum_{i=0}^{2^{t-k-2}-1} (-1)^{|p_{2i+1}(t-k)|+1} \mathcal{N}_s\left(R^{p_{2i+1}^+(t-k)} \ltimes \left\{R_{Q_{2^{k+1}.i+2^k}(t)} \cup R_{Q_{2^{k+1}.i}(t)}\right\}\right)$$
$$= \sum_{i=0}^{2^{t-k-2}-1} (-1)^{|p_{2i+1}(t-k)|+1} \mathcal{N}_s\left(R^{p_{2i+1}^+(t-k)}\right)$$
$$- \sum_{i=0}^{2^{t-k-2}-1} (-1)^{|p_{2i+1}(t-k)|+1} \mathcal{N}_s\left(R^{p_{2i+1}^+(t-k)} \ltimes R_{(1,2,\cdots,k)}\right)$$
$$= \underbrace{\sum_{i=0}^{2^{t-k-2}-1} (-1)^{|p_i(t-k-1)|} \mathcal{N}_s\left(R^{p_i(t-k-1)}\right)}_{\gamma_{\mathcal{P}(t-k-1)}}$$
$$+ \underbrace{\sum_{i=0}^{2^{t-k-2}-1} (-1)^{|p_i(t-k-1)|+1} \mathcal{N}_s\left(R^{p_i(t-k-1)} \ltimes R_{(1,2,\cdots,k)}\right)}_{\sum_{j=1}^{k} \gamma_{\{\mathcal{P}(t-k-1),j\}}}.$$

In the above, the third equation hold by Equation 3.4 and the absorptive law, and the last equation holds by 1) of Lemma 3.13 and the fact $|p_{2i+1}(t-k)| = |p_i(t-k-1)|+1$. Combine the results of the above steps, we have $\gamma_{\mathcal{P}(t)}^k = A + B = \gamma_{\mathcal{P}(t)}^{k+1} + \gamma_{\mathcal{P}(t-k-1)} + \sum_{j=1}^{k} \gamma_{\{\mathcal{P}(t-k-1),j\}}$, the theorem is proved. □

**Corollary 3.15** $\gamma_{\mathcal{P}(t)}^0 = \gamma_{\mathcal{P}(t)}^1 + \gamma_{\mathcal{P}(t-1)}$ for $t \geq 2$.

**Proof** In the proof of Theorem 3.14, note that when $k = 0$, $B + A_2 = \gamma_{\mathcal{P}(t-1)}$. □

**Proof of Theorem 3.3:**

Given an arbitrary binary sequence $s$, we have

$$wt(s \oplus T^t s) - wt(s \oplus T^{t-1} s) = \sum_{i=0}^{2^{t-1}-1} (-1)^{|p_i(t)|+1} \mathcal{N}_s(R^{p_i^+(t)} \ltimes R_{Q_i(t)}) = \gamma^0_{\mathcal{P}(t)}.$$

By theorem 3.14, we have

$$\begin{aligned}
\gamma^0_{\mathcal{P}(t)} &= \gamma^1_{\mathcal{P}(t)} + \gamma_{\mathcal{P}(t-1)} \\
&= \gamma^2_{\mathcal{P}(t)} + \gamma_{\mathcal{P}(t-2)} + \gamma_{\{\mathcal{P}(t-2),1\}} + \gamma_{\mathcal{P}(t-1)} \\
&= \gamma^3_{\mathcal{P}(t)} + \gamma_{\mathcal{P}(t-3)} + \gamma_{\{\mathcal{P}(t-3),1\}} + \gamma_{\{\mathcal{P}(t-3),2\}} + \gamma_{\mathcal{P}(t-2)} + \gamma_{\{\mathcal{P}(t-2),1\}} + \gamma_{\mathcal{P}(t-1)} \\
&= \cdots \cdots .
\end{aligned}$$

Generally, we have

$$\gamma^0_{\mathcal{P}(t)} = \gamma^k_{\mathcal{P}(t)} + \sum_{j=t-k}^{t-1} \gamma_{\mathcal{P}(j)} + \sum_{\ell=1}^{t-2} \sum_{j=t-k}^{t-j-1} \gamma_{\{\mathcal{P}(j),\ell\}}. \tag{3.6}$$

Let $k = t - 1$, we have

$$\begin{aligned}
\gamma^0_{\mathcal{P}(t)} &= \gamma^{t-1}_{\mathcal{P}(t)} + \sum_{j=1}^{t-1} \gamma_{\mathcal{P}(j)} + \sum_{\ell=1}^{t-2} \sum_{j=1}^{t-j-1} \gamma_{\mathcal{P}(j),\ell} \\
&= \gamma - \sum_{i=1}^{t-1} \mathcal{N}_s(R_i) + \sum_{j=1}^{t-1} \gamma_{\mathcal{P}(j)} + \sum_{i=1}^{t-1} \Big(\sum_{j+\ell=i} \gamma_{\mathcal{P}(j),\ell}\Big) \\
&= \gamma + 2 \sum_{i=1}^{t-1} \gamma_{\mathcal{P}(i)} \\
&= \gamma + 2 \sum_{i_1+i_2+\cdots+i_\ell < t} (-1)^\ell \mathcal{N}_s(R_{i_1} R_{i_2} \cdots R_{i_\ell}).
\end{aligned}$$

Theorem 3.3 is proved. □

The sequences in the examples below are take from [5] and known as the *almost perfect autocorrelation sequences*.[1]

**Example 3.16** *Reconsider the sequence in Section 2, $s = (110100000110010101111100)$ with period 24. Write it as*

$$s = (R_2 R_1 R_1 R_6 R_2 R_2 R_1 R_1 R_1 R_1 R_4 R_2).$$

---

[1] An almost perfect autocorrelation sequence is the one such that $C_s(t) \equiv 0$ for all $t \neq 0, N/2$.

One can check that $\gamma = 12$, $\gamma_{\mathcal{P}(0)} = 0$, $\gamma_{\mathcal{P}(1)} = \gamma_{\mathcal{P}(23)} = \mathcal{N}_s(R_1) = 6$, $\gamma_{\mathcal{P}(2)} = \gamma_{\mathcal{P}(22)} = -\mathcal{N}_s(R_2) + \mathcal{N}_s(R_1 R_1) = 4 - 4 = 0$, $\gamma_{\mathcal{P}(3)} = \gamma_{\mathcal{P}(21)} = \cdots = \gamma_{\mathcal{P}(10)} = \gamma_{\mathcal{P}(12)} = 0$, $\gamma_{\mathcal{P}(11)} = \gamma_{\mathcal{P}(13)} = 5$, $\gamma_{\mathcal{P}(12)} = -10$. Hence we have the following table.

**Table I:** *The relation between the autocorrelation function and the run structure of s.*

| $i$ | 0 | 1 | 2 | $\cdots$ | 10 | 11 | 12 | 13 | 14 | $\cdots$ | 22 | 23 |
|---|---|---|---|---|---|---|---|---|---|---|---|---|
| $\gamma_{\mathcal{P}(i)}$ | 0 | -6 | 0 | 0 | 0 | 5 | -10 | 5 | 0 | 0 | 0 | -6 |
| $Wt(s \oplus T^{i+1}s) - Wt(s \oplus T^i s)$ | 12 | 0 | 0 | 0 | 0 | 10 | -10 | 0 | 0 | 0 | 0 | -12 |
| $Wt(s \oplus T^{i+1}s)$ | 12 | 12 | 12 | 12 | 12 | 22 | 12 | 12 | 12 | 12 | 12 | 0 |
| $C_s(i+1)$ | 0 | 0 | 0 | 0 | 0 | -20 | 0 | 0 | 0 | 0 | 0 | 24 |

**Example 3.17** *Consider the almost perfect sequence with period 36,*

$$s = (000101001000111110011010110111000001).$$

*One can check that $\gamma = 18$, $\mathcal{N}_s(R_1) = 9$, $\mathcal{N}_s(R_2) = 4$, $\mathcal{N}_s(R_3) = 3$, $\mathcal{N}_s(R_5) = 2$, $\mathcal{N}_s(R_1 R_1) = 4$, $\mathcal{N}_s(R_1 R_2) = 2$, $\mathcal{N}_s(R_1 R_1 R_1) = 2$, and $\mathcal{N}_s(R_2 R_1) = 3$. Thus, we have $C_s(1) = C_s(2) = C_s(3) = C_s(4) = 0$ by Theorem 3.1. For the other autocorrelation coefficients, one can check them similarly.*

**Example 3.18** *An almost perfect sequence with period 76*

$s = (0011000001000010010111100010111010110001001111101111011010000111010001010011)$

*One can check that $\gamma = 38$, $\mathcal{N}_s(R_1) = 19$, $\mathcal{N}_s(R_2) = 8$, $\mathcal{N}_s(R_3) = 5$, $\mathcal{N}_s(R_4) = 4$, $\mathcal{N}_s(R_5) = 2$, $\mathcal{N}_s(R_1 R_1) = 8$, $\mathcal{N}_s(R_1 R_1 R_1) = 2$, $\mathcal{N}_s(R_2 R_1) = 2$, and $\mathcal{N}_s(R_1 R_2) = 5$. Thus, by Theorem 3.3, $C_s(1) = C_s(2) = C_s(3) = C_s(4) = 0$. The other autocorrelation coefficients can be obtained similarly.*

## 4 Examples

In the following, we directly calculate $wt(s \oplus T^t s) - wt(s \oplus T^{t-1}s)$ for the case of $t = 3$ and $t = 4$.

**Example 4.1** *Let s be a binary sequence with period N and the total number of runs $\gamma$. Then,*

$$wt(s \oplus T^3 s) - wt(s \oplus T^2 s) = \gamma - 2\mathcal{N}_s(R_1) - 2\mathcal{N}_s(R_2) + 2\mathcal{N}_s(R_1 R_1).$$

**Proof** Firstly, by Definition 3.4, one can calculate $Q_i(t)$ for $t = 3$ as the following table shown.

**Table II:** The values of $Q_i(t)$ for $t = 3$.

| $i$ | 0 | 1 | 2 | 3 |
|---|---|---|---|---|
| $p_i(3)$ | (3) | (1,2) | (2,1) | (1,1,1) |
| $Q_i(3)$ | $3, 4, 5, \cdots$ | $1, 2$ | $2$ | $1, 3, 4, 5 \cdots$ |

By the absorptive law, we have $\sum_{j \geq 1} \mathcal{N}_s(R_j R_2) = \mathcal{N}_s(R_2)$, and $\sum_{j \geq 1} \mathcal{N}_s(R_j R_1 R_1) = \mathcal{N}_s(R_1 R_1)$. Thus,

$$wt(s \oplus sT^3 s) - wt(s \oplus sT^2 s) = \sum_{j \geq 3, \ell \geq 3} \mathcal{N}_s(R_j R_\ell) - \mathcal{N}_s(R_2 R_1) - \mathcal{N}_s(R_2 R_2)$$
$$- \sum_{j \geq 2} \mathcal{N}_s(R_j R_1 R_2) + \mathcal{N}_s(R_1 R_1 R_1) + \sum_{\ell \geq 3} \mathcal{N}_s(R_1 R_1 R_\ell). \quad (4.1)$$

Since

1. $\sum_{j \geq 3, \ell \geq 3} \mathcal{N}_s(R_j R_\ell) = \sum_{j \geq 2, \ell \geq 3} \mathcal{N}_s(R_j R_\ell) - \sum_{\ell \geq 3} \mathcal{N}_s(R_2 R_\ell)$;

2. $\sum_{j \geq 2} \mathcal{N}_s(R_j R_1 R_2) = \mathcal{N}_s(R_1 R_2) - \mathcal{N}_s(R_1 R_1 R_2)$,

we have

$$wt(s \oplus sT^3 s) - wt(s \oplus sT^2 s)$$
$$= - \left( \mathcal{N}_s(R_2 R_1) + \mathcal{N}_s(R_2 R_2) + \sum_{\ell \geq 3} \mathcal{N}_s(R_2 R_\ell) \right) + \sum_{j \geq 2, \ell \geq 3} \mathcal{N}_s(R_j R_\ell)$$
$$+ \left( \mathcal{N}_s(R_1 R_1 R_1) + \mathcal{N}_s(R_1 R_1 R_2) + \sum_{\ell \geq 3} \mathcal{N}_s(R_1 R_1 R_\ell) \right) - \mathcal{N}_s(R_1 R_2)$$
$$= - \mathcal{N}_s(R_2) + \mathcal{N}_s(R_1 R_1) - \mathcal{N}_s(R_1 R_2) + \sum_{j \geq 2, \ell \geq 3} \mathcal{N}_s(R_j R_\ell)$$

Since

$$\sum_{j \geq 2, \ell \geq 3} \mathcal{N}_s(R_j R_\ell) = \sum_{\ell \geq 3} \mathcal{N}_s(R_\ell) - \sum_{\ell \geq 3} \mathcal{N}_s(R_1 R_\ell),$$

we have

$$wt(s \oplus sT^3 s) - wt(s \oplus sT^2 s)$$
$$= - \mathcal{N}_s(R_2) + 2\mathcal{N}_s(R_1 R_1) - \left( \mathcal{N}_s(R_1 R_1) + \mathcal{N}_s(R_1 R_2) + \sum_{\ell \geq 3} \mathcal{N}_s(R_1 R_\ell) \right) + \sum_{\ell \geq 3} \mathcal{N}_s(R_\ell)$$
$$= - \mathcal{N}_s(R_2) + 2\mathcal{N}_s(R_1 R_1) - \mathcal{N}_s(R_1) + \sum_{\ell \geq 3} \mathcal{N}_s(R_\ell)$$
$$= - \mathcal{N}_s(R_2) + 2\mathcal{N}_s(R_1 R_1) - \mathcal{N}_s(R_1) + (\gamma - \mathcal{N}_s(R_1) - \mathcal{N}_s(R_2))$$
$$= \gamma - 2\mathcal{N}_s(R_1) - 2\mathcal{N}_s(R_2) + 2\mathcal{N}_s(R_1 R_1),$$

which completes the proof. $\square$

**Example 4.2** *Let $s$ be a binary sequence with period $N$ and the total number of runs $\gamma$. Then,*

$$wt(s \oplus sT^4s) - wt(s \oplus sT^3s)$$
$$= \gamma - 2\sum_{i=1}^{3}\mathcal{N}_s(R_i) + 2\left(\mathcal{N}_s(R_1R_1) + \mathcal{N}_s(R_1R_2) + \mathcal{N}_s(R_2R_1)\right) - 2\mathcal{N}_s(R_1R_1R_1).$$

**Proof** Firstly, one can check $Q_i(t)$ for $t = 4$ as the following table shown.

**Table III:** The values of $Q_i(t)$ for $t = 4$.

| $i$ | 0 | 1 | 2 | 3 | 4 | 5 | 6 | 7 |
|---|---|---|---|---|---|---|---|---|
| $p_i(4)$ | (4) | (1,3) | (2,2) | (1,1,2) | (3,1) | (1,2,1) | (2,1,1) | (1,1,1,1) |
| $Q_i(4)$ | $4,5,\cdots$ | $1,2,3$ | $2,3$ | $1,4,5,\cdots$ | $3$ | $1,2,4,5,\cdots$ | $2,4,5,\cdots$ | $1,3$ |

Thus, $wt(s \oplus T^4s) - wt(s \oplus T^3s)$ is equal to:

$$\sum_{j,\ell \geq 4}\mathcal{N}_s(R_jR_\ell) - \mathcal{N}_s(R_3R_1) - \mathcal{N}_s(R_3R_2) - \mathcal{N}_s(R_3R_3)$$
$$- \sum_{j\geq 2}\mathcal{N}_s(R_jR_2R_2) - \sum_{j\geq 2}\mathcal{N}_s(R_jR_2R_3) + \mathcal{N}_s(R_1R_2R_1) + \sum_{\ell\geq 4}\mathcal{N}_s(R_1R_2R_\ell)$$
$$- \sum_{j\geq 3}\mathcal{N}_s(R_jR_1R_3) + \mathcal{N}_s(R_2R_1R_1) + \mathcal{N}_s(R_2R_1R_2) + \sum_{\ell\geq 4}\mathcal{N}_s(R_2R_1R_\ell)$$
$$+ \sum_{j\geq 2}\mathcal{N}_s(R_jR_1R_1R_2) + \sum_{j\geq 2,\ell\geq 4}\mathcal{N}_s(R_jR_1R_1R_\ell) - \mathcal{N}_s(R_1R_1R_1R_1) - \mathcal{N}_s(R_1R_1R_1R_3).$$

Note that the absorptive law has been used in the above arguments. Since

$$\sum_{j,\ell\geq 4}\mathcal{N}_s(R_jR_\ell) = \sum_{j\geq 3,\ell\geq 4}\mathcal{N}_s(R_jR_\ell) - \sum_{\ell\geq 4}\mathcal{N}_s(R_3R_\ell);$$

$$\sum_{j\geq 2}\mathcal{N}_s(R_jR_2R_2) = \mathcal{N}_s(R_2R_2) - \mathcal{N}_s(R_1R_2R_2);$$

$$\sum_{j\geq 2}\mathcal{N}_s(R_jR_2R_3) = \mathcal{N}_s(R_2R_3) - \mathcal{N}_s(R_1R_2R_3);$$

$$\sum_{j\geq 3}\mathcal{N}_s(R_jR_1R_3) = \sum_{j\geq 2}\mathcal{N}_s(R_jR_1R_3) - \mathcal{N}_s(R_2R_1R_3);$$

$$\sum_{j\geq 2}\mathcal{N}_s(R_jR_1R_1R_2) = \mathcal{N}_s(R_1R_1R_2) - \mathcal{N}_s(R_1R_1R_1R_2);$$

$$\sum_{j\geq 2,\ell\geq 4}\mathcal{N}_s(R_jR_1R_1R_\ell) = \sum_{\ell\geq 4}\mathcal{N}_s(R_1R_1R_\ell) - \sum_{\ell\geq 4}\mathcal{N}_s(R_1R_1R_1R_\ell),$$

using the absorptive law again, we have

$$
\begin{aligned}
&wt(s \oplus T^4 s) - wt(s \oplus T^3 s) \\
&= -\mathcal{N}_s(R_3) + \mathcal{N}_s(R_1 R_2) + \mathcal{N}_s(R_2 R_1) - \mathcal{N}_s(R_1 R_1 R_1) \\
&\quad + \sum_{j \geq 3, \ell \geq 4} \mathcal{N}_s(R_j R_\ell) - \mathcal{N}_s(R_2 R_2) - \mathcal{N}_s(R_2 R_3) \\
&\quad - \sum_{j \geq 2} \mathcal{N}_s(R_j R_1 R_3) + \mathcal{N}_s(R_1 R_1 R_2) + \sum_{\ell \geq 4} \mathcal{N}_s(R_1 R_1 R_\ell).
\end{aligned}
$$

Since

$$
\sum_{j \geq 3, \ell \geq 4} \mathcal{N}_s(R_j R_\ell) = \sum_{j \geq 2, \ell \geq 4} \mathcal{N}_s(R_j R_\ell) - \sum_{\ell \geq 4} \mathcal{N}_s(R_2 R_\ell);
$$

$$
\sum_{j \geq 2} \mathcal{N}_s(R_j R_1 R_3) = \mathcal{N}_s(R_1 R_3) - \mathcal{N}_s(R_1 R_1 R_3),
$$

using the absorptive law again, we have

$$
\begin{aligned}
&wt(s \oplus T^4 s) - wt(s \oplus T^3 s) \\
&= -\mathcal{N}_s(R_3) + \mathcal{N}_s(R_1 R_2) + \mathcal{N}_s(R_2 R_1) - \mathcal{N}_s(R_1 R_1 R_1) \\
&\quad + \sum_{j \geq 2, \ell \geq 4} \mathcal{N}_s(R_j R_\ell) - \mathcal{N}_s(R_2) + \mathcal{N}_s(R_2 R_1) \\
&\quad - \mathcal{N}_s(R_1 R_3) + \mathcal{N}_s(R_1 R_1) - \mathcal{N}_s(R_1 R_1 R_1).
\end{aligned}
$$

Since

$$
\sum_{j \geq 2, \ell \geq 4} \mathcal{N}_s(R_j R_\ell) = \sum_{\ell \geq 4} \mathcal{N}_s(R_\ell) - \sum_{\ell \geq 4} \mathcal{N}_s(R_1 R_\ell),
$$

we have

$$
\begin{aligned}
&wt(s \oplus T^4 s) - wt(s \oplus T^3 s) \\
&= -\mathcal{N}_s(R_3) + \mathcal{N}_s(R_1 R_2) + \mathcal{N}_s(R_2 R_1) - \mathcal{N}_s(R_1 R_1 R_1) \\
&\quad - \mathcal{N}_s(R_2) + \mathcal{N}_s(R_1 R_1) + \mathcal{N}_s(R_2 R_1) - \mathcal{N}_s(R_1 R_1 R_1) + \sum_{\ell \geq 4} \mathcal{N}_s(R_\ell) \\
&\quad - \sum_{\ell \geq 4} \mathcal{N}_s(R_1 R_\ell) - \mathcal{N}_s(R_1 R_3) \\
&= -\mathcal{N}_s(R_3) + \mathcal{N}_s(R_1 R_2) + \mathcal{N}_s(R_2 R_1) - \mathcal{N}_s(R_1 R_1 R_1) \\
&\quad - \mathcal{N}_s(R_2) + \mathcal{N}_s(R_1 R_1) + \mathcal{N}_s(R_2 R_1) - \mathcal{N}_s(R_1 R_1 R_1) + \sum_{\ell \geq 4} \mathcal{N}_s(R_\ell) \\
&\quad - \mathcal{N}_s(R_1) + \mathcal{N}_s(R_1 R_1) + \mathcal{N}_s(R_1 R_2).
\end{aligned}
$$

Thus, we have

$$wt(s \oplus T^4 s) - wt(s \oplus T^3 s)$$
$$= \gamma - 2\sum_{i=1}^{3}\mathcal{N}_s(R_i) + 2\left(\mathcal{N}_s(R_1R_1) + \mathcal{N}_s(R_1R_2) + \mathcal{N}_s(R_2R_1)\right) - 2\mathcal{N}_s(R_1R_1R_1).$$

The result is proved. □

## 5 Applications

In this section, we give some applications of the theoretic result, which includes the ZCZ sequence, the cyclic difference sets, and the circulant Hadamard Matrices.

### 5.1 ZCZ Sequences

The *zero autocorrelation zone* ($Z_{ACZ}$) of a binary sequence $s$ is defined as:

$$Z_{ACZ} = \max\{T : C_s(w) = 0, \forall w \neq 0, |w| \leq T\}.$$

Zero autocorrelation zone sequences (ZCZ sequences) are useful in the approximately synchronized CDMA (AS-CDMA) communication systems [6]. A new characterization of ZCZ sequences can be given according to Theorem 3.3.

**Theorem 5.1** *The following items characterizes a ZCZ sequence with $Z_{ACZ} = D$.*

1. $\gamma = N/2$;
2. $\mathcal{N}_s(R_1) = \gamma/2$;
3. $\gamma_{\mathcal{P}(t)} = 0$, *for $t = 2, 3, \cdots, D-1$.*

As an example, in the following, we discuss the sequence with $Z_{ACZ} \leq 4$ and give all such ZCZ sequences with period 12. Let $P_i$ denote a string of consecutive $R_i$ flanked at either end by the runs different from $R_i$. Let $\mathcal{L}(P_i)$ be the number of $R_i$s in $P_i$, and $\mathcal{N}_s(P_i)$ be the number of $P_i$s in $s$. By counting $\mathcal{N}_s(R_i) - \mathcal{N}_s(R_iR_i)$ and $\mathcal{N}_s(R_i) - \mathcal{N}_s(R_iR_iR_i)$ for each $P_i$, one can have,

1. $\mathcal{N}_s(R_iR_i) = \mathcal{N}_s(R_i) - \mathcal{N}_s(P_i)$;
2. $\mathcal{N}_s(R_iR_iR_i) = \mathcal{N}_s(R_i) - 2\mathcal{N}_s(P_i) + \sum_{\mathcal{L}(P_i)=1}\mathcal{N}_s(P_i)$.

Then, combine the cases $t = 2, 3$ of Theorem 3.3, we have

**Proposition 5.2** $C_s(1) = C_s(3) \iff \sum_{j \geq 3}\mathcal{N}_s(R_j) = \mathcal{N}_s(P_1).$

The following items characterize a sequence with $Z_{ACZ} \leq 4$.

**Proposition 5.3** *A sequence with $Z_{ACZ} \leq 4$ is decided by the following items.*

1. $\gamma = N/2$ ($C_s(1) = 0$);

2. $\mathcal{N}_s(R_1) = \gamma/2$ ($C_s(1) = C_s(2)$);

3. $\mathcal{N}_s(R_2) + \mathcal{N}_s(P_1) = \gamma/2$ ($C_s(2) = C_s(3)$);

4. $\sum_{\ell \geq 4} \mathcal{N}_s(R_\ell) + \mathcal{N}_s(P_1) - \sum_{\mathcal{L}(P_1)=1} \mathcal{N}_s(P_1) + (\mathcal{N}_s(R_2 R_1) + \mathcal{N}_s(R_1 R_2)) = \gamma/2$ ($C_s(3) = C_s(4)$).

Now, we can give all the sequences with $Z_{ACZ} \leq 4$ of period 12.

**Proposition 5.4** *There are totally 20 sequences with $Z_{ACZ} \leq 4$ of period 12.*

**Proof** For the sequence with $Z_{ACZ} \leq 4$ of period 12, one can have that $\gamma = 6$ and $\mathcal{N}_s(R_1) = 3$. We distinguish 3 cases to discussion.

1. $\mathcal{N}_s(P_1) = 1$. In this case, $\sum_{j \geq 3} \mathcal{N}_s(R_j) = 1$. Hence, $s$ is composed of 6 runs, namely, $R_1, R_1, R_1, R_2, R_2,$ and $R_5$. Since $\mathcal{N}_s(R_1 R_2) + \mathcal{N}_s(R_2 R_1) = 1$. $s$ is with the form:

$$(R_1 R_1 R_1 R_2 R_2 R_5), (R_1 R_1 R_1 R_5 R_2 R_2).$$

2. $\mathcal{N}_s(P_1) = 2$. In this case, $\sum_{j \geq 3} \mathcal{N}_s(R_j) = 2$. Hens, $s$ is composed of 6 runs, namely, $R_1, R_1, R_1, R_2, R_3,$ and $R_4$. Since $\mathcal{N}_s(R_1 R_2) + \mathcal{N}_s(R_2 R_1) = 1$. $s$ is with the form:

$$(R_1 R_3 R_1 R_1 R_2 R_4), (R_1 R_3 R_1 R_1 R_4 R_2), (R_1 R_4 R_1 R_1 R_2 R_3), (R_1 R_4 R_1 R_1 R_3 R_2),$$

$$(R_1 R_3 R_2 R_1 R_1 R_4), (R_1 R_2 R_3 R_1 R_1 R_4), (R_1 R_4 R_2 R_1 R_1 R_3), (R_1 R_2 R_4 R_1 R_1 R_3).$$

3. $\mathcal{N}_s(P_1) = 3$. In this case, $\sum_{j \geq 3} \mathcal{N}_s(R_j) = 3$, which yields $s = (1000)$ or $(0111)$. Thus, no such sequence exists.

Note that each of the above forms associates with 2 sequences, the result follows. □

## 5.2 Cyclic Difference Sets and Circulant Hadamard Matrices

A $(v, k, \lambda)$-*cyclic difference set* is a subset $D = \{d_0, d_1, \cdots d_{k-1}\}$ of a cyclic group $(G, +)$ of order $v$ such that for each $g \neq 0 \in G$, there exist exactly $\lambda$ ordered pairs $i, j$ with $i \neq j$ satisfies $d_i - d_j = g$. It is well known that a $(v, k, \lambda)$-cyclic difference set is equivalent to a binary sequence with period $v$ and constant out-of-phase autocorrelation coefficients $v - 4(k - \lambda)$ [3]. Hence, by Theorem 3.3, we have

**Theorem 5.5** *A cyclic difference set of G is equivalent to a $|G|$-period binary sequence such that:*

1. $\mathcal{N}_s(R_1) = \gamma/2$;

2. $\gamma_{\mathcal{P}(t)} = 0$, for $t = 2, 3, \cdots, |G| - 2$,

A *circulant Hadamard matrix* is a $N \times N$ matrix $H$ with entries $\pm 1$ of the form

$$\begin{pmatrix} v_0 & v_1 & \cdots & v_{N-1} \\ v_{N-1} & v_0 & \cdots & v_{N-2} \\ \cdots & \cdots & \cdots & \cdots \\ v_1 & v_2 & \cdots & v_0 \end{pmatrix}$$

such that any two rows of $H$ are orthogonal. By letting $s_i = (v_i + 1)/2$, $H$ corresponds to a binary sequence $s = (s_0, s_1, \cdots, s_{N-1})$ with $C_s(1) = C_s(2) = \cdots = C_s(N-1) = 0$ and vice versa. Hence, we have

**Theorem 5.6** *A circulant Hadamard matrix of order $N$ is equivalent to an $N$-period binary sequence of*

1. $\gamma = N/2$;

2. $\mathcal{N}_s(R_1) = \gamma/2$;

3. $\gamma_{\mathcal{P}(t)} = 0$, for $t = 2, 3, \cdots, N - 2$.

The prominent *Circulant Hadamard Matrix Conjecture* [7] asserts that there is no circulant Hadamard matrix besides $N = 1, 4$. The above results provide a new approach to analyze the conjecture, which can be seen from the following simple examples.

**Example 5.7** *Let $H$ be an circulant Hadamard matrix with order $N \neq 4$ and $s$ be the binary sequence corresponding to $H$. Then $\mathcal{N}_s(R_2) \neq 0$.*

**Proof** By Theorem 5.6, $\mathcal{N}_s(R_1) = \gamma/2 = N/4$. If $\mathcal{N}_s(R_2) = 0$, then by item (3) of Proposition 5.3, $\mathcal{N}_s(P_1) = N/4$, which yields $\mathcal{N}_s(R_j) = \gamma/2$, i.e., $s = (1000)$ or $(1110)$, a contradiction. □

**Example 5.8** *There exists no circulant Hadamard matrix with order 8.*

**Proof** Let $H$ be a circulant Hadamard matrix with order $N = 8$. By Proposition 5.3, $\gamma = 4$ and $\mathcal{N}_s(R_1) = 2$ and $\sum_{j \geq 2} \mathcal{N}_s(R_j) = 2$. By Example 5.7, $\mathcal{N}_s(R_2) \neq 0$, as yields $\mathcal{N}_s(R_2) = \mathcal{N}_s(R_4) = 1$. By Proposition 5.2, $s$ is with the form $(R_1, R_1, R_2, R_4)$ or $(R_1, R_1, R_4, R_2)$. Hence $2 = \frac{1}{2}\gamma = \sum_{\ell \geq 4} \mathcal{N}_s(R_\ell) + \mathcal{N}_s(P_1) - \sum_{\mathcal{L}(P_1)=1} \mathcal{N}_s(P_1) + (\mathcal{N}_s(R_2 R_1) + \mathcal{N}_s(R_1 R_2)) = 1 + 1 - 0 + 1 = 3$, a contradiction. □

# Acknowledgement

The author would like to thank Prof. Rongquan Feng for his guidance and encouragement.

# References

[1] S. W.Golomb, *Shift Register Sequences*, Holden-Day, San Francisoc, California, 1967.

[2] P. Fan and M. Darnell, *Sequence Design for Communications Applications*, Research Studies Press, London, 1996.

[3] S. W. Golomb and G. Gong, *Signal designs with good correlation: for wireless communications, cryptography and radar applications.* Cambridge, U.K.: Cambridge University Press, 2005.

[4] G. E.Andrews, *The Theory of Partitions. Cambridge,* England: Cambridge University Press, 1998.

[5] J. Wolfmann, *Almost Perfect Autocorrelation Sequences*, IEEE Trans. Inf. Theory, vol. 38, no. 4, pp. 1412-1418, Jul. 1992.

[6] X. Deng and P. Fan, *Spreading sequence sets with zero correlation zone*, Electronics Letters 36, May 2000, 993-994.

[7] B. Schmidt, *Cyclotomic Ingegers and Finite Geometry*, Journal of the American Mathematical Society, vol. 12, no. 4, pp. 929-952, May 1999.